\documentclass[iop]{emulateapj}
\usepackage{amsmath}
\usepackage{natbib}
\begin{document}

\title{Gas Sloshing in Abell\,2204: Constraining the Properties of the Magnetized Intracluster Medium}
\author{Huanqing Chen\altaffilmark{1,2,3,*}}
\author{Christine Jones\altaffilmark{2}}
\author{Felipe Andrade-Santos\altaffilmark{2}}
\author{John A. ZuHone\altaffilmark{2}}
\author{Zhiyuan Li\altaffilmark{1,3}}
\altaffiltext{1}{School of Astronomy and Space Science, Nanjing University, 163 Xianlin Ave., Nanjing, Jiangsu 210046, China; lizy@nju.edu.cn}
\altaffiltext{2}{Harvard-Smithsonian Center for Astrophysics, 60 Garden St., Cambridge, MA 02138, USA}
\altaffiltext{3}{Key Laboratory of Modern Astronomy and Astrophysics (Nanjing University), Ministry of Education, Nanjing 210046, China}
\altaffiltext{*}{Current address: Department of Astronomy and Astrophysics, University of Chicago,  5640 S. Ellis Ave, Chicago, IL 60637, USA; \email{hqchen@uchicago.edu}}
\begin{abstract}

The rich galaxy cluster Abell\,2204 exhibits edges in its X-ray surface brightness at $\sim 65$ and $35 {\rm~ kpc}$ west and east of its center, respectively.
The presence of these edges, which were interpreted as sloshing cold fronts, implies that the intracluster medium was recently disturbed. 
We analyze the properties of the intracluster medium using multiple {\it Chandra} observations of Abell\,2204. 
We find a density ratio $n_{\rm in}/n_{\rm out} = 2.05\pm0.05$ and a temperature ratio $T_{\rm out}/T_{\rm in} = 1.91\pm0.27$ (projected, or $1.87\pm0.56$ deprojected) across the western edge, and correspondingly $n_{\rm in}/n_{\rm out} =  1.96\pm0.05$ and $T_{\rm out}/T_{\rm in} =1.45\pm0.15$ (projected, or $1.25\pm0.26$ deprojected) across the eastern edge. 
These values are typical of cold fronts in galaxy clusters. 
This, together with the spiral pattern observed in the cluster core, supports the sloshing scenario for Abell\,2204.  
 No Kelvin-Helmholtz eddies are observed along the cold front surfaces,
indicating that they are effectively suppressed by some physical mechanism.
We argue that the suppression is likely facilitated by the magnetic fields amplified in the sloshing motion, and deduce from the measured gas properties that the magnetic field strength should be greater than $24\pm6$ $\mu$G and $32\pm8$ $\mu$G along the west and east cold fronts, respectively. 
\end{abstract}

\section{introduction}
Galaxy clusters are an ideal landscape for tracing structural growth in the hierarchical universe.
One of the initial surprises brought by the {\em Chandra X-ray Observatory} was the discovery of ``cold fronts'' (CFs) in galaxy clusters.
First found in Abell\,2142 \citep{markevitch2000} and Abell\,3667 \citep{vikhlinin2001},
CFs manifest themselves as bright, arc-shaped surface brightness edges. 
Most CFs are thought to be the product of gas ``sloshing'', which occurs following a non-head-on encounter between two clusters \citep{markevitch2001,ascasibar2006,markevitch2007}. 
When the subcluster approaches the main cluster, the latter's intracluster medium (ICM) becomes detached from the dark matter (DM) halo under the influence of ram pressure.
Later on, the ICM falls back into the potential well of the main cluster and starts sloshing.
Consequently, the cool, dense gas in the core is driven outward, where it is held by the hotter, less dense gas, forming a contact discontinuity.
Observationally, gas temperature inside the discontinuity is lower than in the outside (hence the name of {\it cold front}),
whereas gas density drops abruptly, maintaining a nearly continuous gas pressure across the CF. 
These characteristics are markedly different from those of shock fronts. 

Numerical simulations \citep[e.g.,][]{ascasibar2006} suggest that gas sloshing can be easily induced by minor mergers, and that the sloshing features may remain detectable for more than a billion years,
which naturally explains why CFs are so commonly found in nearby galaxy clusters.
The minor merger scenario is further supported by the fact that many clusters hosting CFs exhibit a spiral pattern
in the core \citep[e.g.,][]{johnson2010,johnson2012,nulsen2013,santos2016}, which
is also seen in high-fidelity gas sloshing simulations \citep{ascasibar2006,zuhone2011,roediger2013,zuhone2015}  	
and can be explained as the result of the angular momentum being transported from the subcluster to
the main cluster. 

The CFs are excellent tools for studying cluster dynamics, as well as the physics of the ICM \citep{markevitch2007,zuhone2016}.
For instance, the relatively intact cold fronts, not easily deformed by Rayleigh-Taylor 
instability, can place constraints on the mass of the underlying DM halo in clusters \citep{vikhlinin2002}. 
Observations also show that many CFs have very sharp edges,  with widths less than $\sim 5~\rm kpc$ (e.g., \citealp{vikhlinin2001b,sanders2016,werner2016}).
\citet{vikhlinin2001} found that the CFs in the merging cluster Abell\,3667 show
no sign of Kelvin-Helmholtz instability (KHI) in the direction of merging. 
They proposed that KHI is suppressed by magnetic fields ``frozen'' along the surface of the CFs.
The estimated field strength based on the suppression of KHI is on the order of 10~$\mu$G, 
and the magnetic pressure is approximately 0.1--0.2 of the thermal pressure.
KHI at sloshing CFs has been extensively studied in simulations which take various factors into consideration, such as viscosity and magnetic fields \citep{zuhone2011,zuhone2015,zuhone2016,roediger2013}. 
In particular, simulations with magnetic fields find that highly magnetized layers (compared to other parts of the ICM) will easily form underneath the CFs due to shear amplification by the sloshing motions. The magnetic pressure in these layers can be nearly in equipartition with the thermal pressure. This relatively strong magnetic field may play a crucial role in suppressing KHI. Viscosity may play a similar role.

In this work, we present an X-ray study of the cold fronts in the massive galaxy cluster Abell\,2204 (hereafter A2204).
Early observations by {\it ROSAT} showed that A2204 has a regular X-ray morphology \citep{buote1996}.
It was with {\em Chandra} observations that substructures in the core of A2204 were found \citep{sanders2005,sanders2009}, which include a spiral pattern and two CFs. 
The CFs appear smooth and sharp, showing no sign of KHI \citep{sanders2005}.
Here we use deeper {\em Chandra} exposures, along with new insights from numerical simulations, to analyze these two CFs in detail, in order to constrain the physical properties of the ICM in A2204. 

The remainder of this paper is organized as follows.
In Section~2 we describe the observations and data preparation. 
In Section~3 we analyze the properties of the ICM, focusing on the two CFs in the cluster core.
Section~4 is devoted to understanding the suppression of KHI, and a summary is provided in Section~5.
Throughout this work, we adopt a redshift of z = 0.152 for A2204 \citep{pimbblet2006} and assume a cosmology with $H_0=70 \rm~km~s^{-1}~Mpc^{-1}$, $\Omega_{\Lambda}=0.7$ and $\Omega_{M}=0.3$. 
At the distance of A2204, an arcsecond corresponds to 2.64 kpc.
We measure position angles counterclockwise from the west ($0\degr$).
Quoted errors are at the 1$\sigma$ confidence level, unless otherwise stated.

\section{Data Preparation}

\begin{deluxetable*}{lccccr}
\tablecaption{Log of {\em Chandra} observations\label{tb.1}}
\tablehead{\colhead{ObsID}&\colhead{Obs Date}&\colhead{Exp (ks)}&\colhead{Instrument}&\colhead{Target}&\colhead{PI}}\\
\startdata
 499  & 2000  -07-29 & 10.1& ACIS-S &	Center & A. Fabian\\
 6104 & 2004  -09-20 &  9.6& ACIS-I &	Center 	   & S. Allen\\
 7940 & 2007  -06-06 & 77.1& ACIS-I &	Center 	   & J. Sanders\\
 12898& 2011  -01-13 &  5.0& ACIS-I &	Field (SW) & E. Miller\\
 12897& 2011  -06-06 &  5.0& ACIS-I &	Field (SE) & E. Miller\\
 12896& 2011  -07-16 &  5.0& ACIS-I &	Field (NW) & E. Miller\\
 12895& 2012  -01-12 &  5.0& ACIS-I &	Field (NE) & E. Miller
\enddata
\end{deluxetable*}

In this work, we utilize seven {\it Chandra} Advanced CCD Imaging Spectrometer (ACIS) observations of A2204 (Table \ref{tb.1}). 
Among these, three observations focus on the inner region, which enable a detailed study of the disturbed core. 
The other four are pointed towards the cluster outskirts, which are useful to constrain the large-scale structure of the ICM.
We reprocessed the data following the procedure of \citet{vikhlinin2005}, using the software CHAV and calibration data CALDB v4.6.7.
We examined the light curves and found a rather quiescent instrumental background in all seven observations, thus no
time interval was rejected. 

\begin{figure}[hbt!]
\centerline{
\includegraphics[width=0.5\textwidth]{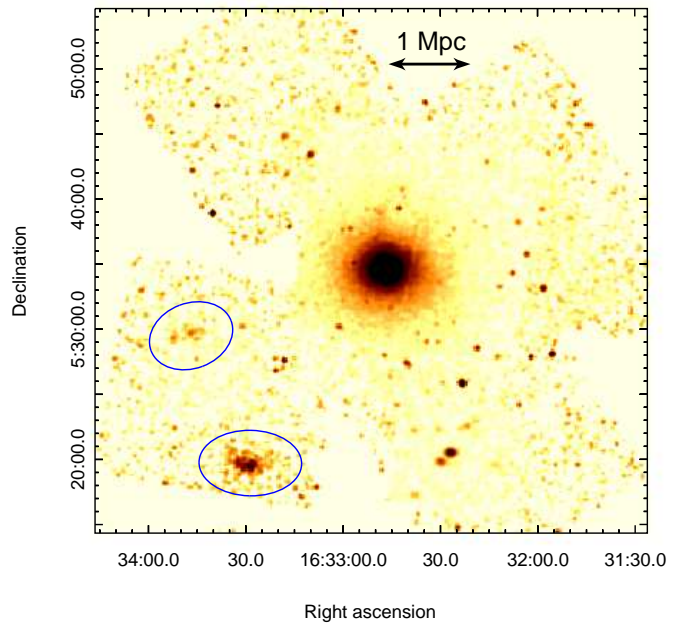}
}
\caption{
Background subtracted, exposure corrected {\em Chandra} 0.5-2 keV image of Abell 2204. A binning of $\approx 8''\times8''$ is adopted.
Two extended sources in the east and southeast are highlighted by ellipses (see discussions of their possible nature in the Appendix).
The two sources in the southwest are most likely discrete sources that appear extended, because they are located far off-axis in ObsID 7940. In ObsID 12898, both sources look point-like.
}
\label{fig.bin16}
\end{figure}

We created background-subtracted and exposure-corrected flux images in the energy bands of 0.5-2.0 and 0.5-7.0 keV, following the procedure of \citet{vikhlinin2005}. 
We have adopted the ``blank sky" background, the level of which was scaled according to the counts 
in the 9.5-12 keV band. The readout background was also statistically subtracted. 
We have merged all seven observations to maximize the signal to noise ratio (S/N) and the field-of-view, as shown in Figure~\ref{fig.bin16}. 
Point sources were identified from this combined image and excluded from subsequent analysis of the diffuse emission.

\begin{figure}[hbt!]
\centerline{
\includegraphics[width=0.5\textwidth]{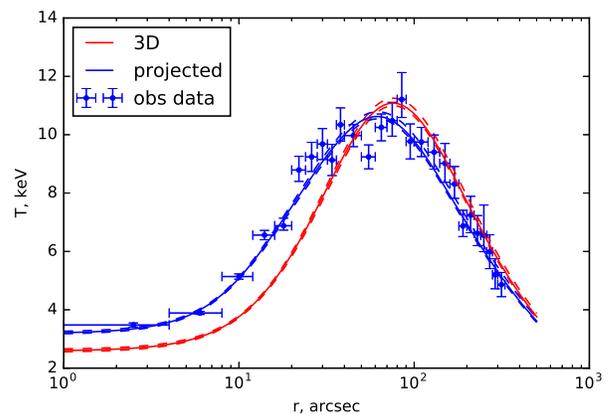}
}
\caption{
Temperature profiles of Abell 2204. The blue data points are derived from spectral fitting. 
A deprojected temperature profile that fits the data is shown as the red curve, whereas
the blue curve is the same model profile in projection. \label{temperature}
}
\end{figure}

\begin{figure*}
\centerline{
\includegraphics[width=\textwidth]{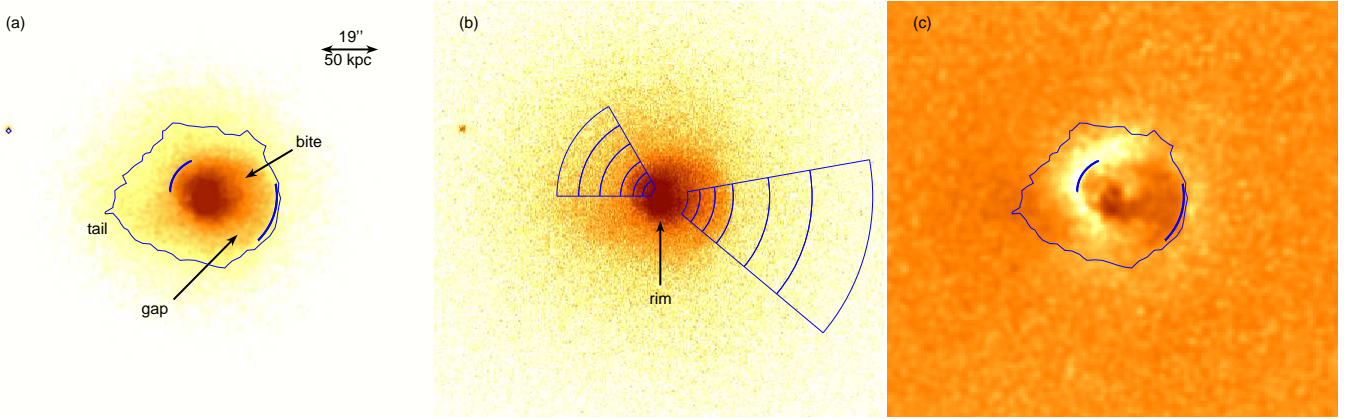}
}
\caption{
(a) Gaussian-smoothed 0.5--7.0 keV flux image of the inner cluster region, with an X-ray intensity contour overlaid.
The blue arcs mark the two surface brightness edges.
(b) Same as in (a), except that no smoothing is applied.
(c) Residual image obtained by subtracting a $\beta$-model image (Section~\ref{sec.discore}).  \label{fig.xraycore}
}
\end{figure*}

For spectral analysis, spectra were extracted from each observation and then co-added to maximize the S/N.  
To obtain an accurate measurement of the gas temperature, the spectral extraction regions were required to contain a minimum of $\sim$4000 total counts. 
We fit the spectra with an absorbed optically-thin plasma model, wabs*apec in XSPEC (version 12.8.2q). 
The column density $N_{\rm H}$ was fixed at $5.67\times10^{20}~\rm~cm^{-2}$, according to the LAB Survey of Galactic HI \citep{kalberla2005}. The redshift was also fixed.

\section{Sloshing cold fronts in A2204}
Figure \ref{fig.bin16} shows the merged {\it Chandra} image of A2204 in the 0.5-2.0 keV band. On large scales up to 1 Mpc, A2204 exhibits a rather relaxed morphology. Two extended sources can be seen at the cluster outskirts, the nature of which are discussed in the Appendix. 
By fitting spectra extracted from concentric annuli, we obtained the temperature profile of A2204, which is shown in Figure~\ref{temperature}.
We confirm that A2204 has a cool core \citep{sanders2005}. 
The projected temperature rises from $\sim$ 3.5 keV at the very center to $\sim$ 11 keV at $80^{\prime\prime}$ ($\sim$ 210 kpc), then drops further out \citep[see also][]{reiprich2009}, consistent with a massive cluster. 
We also employed the deprojection method of \citet{vikhlinin2006}, which results in a deprojected temperature profile (red curve in Figure~\ref{temperature}) similar to the observed profile.

\subsection{Morphology of the disturbed core}\label{sec.discore}

Although A2204 has a regular morphology on large-scale, the cluster is highly disturbed in the core (Figure~\ref{fig.xraycore}), which looks like a tapered water drop with a ``tail'' to the east. 
A spiral pattern is prominent in the core, which starts from the southeast at a radius of $10''$, then wraps outward clockwise to southwest,
reaching a radius of $25''$, and grows fuzzy in the south.
Careful examination reveals a gap in the southwest, and a small ``bite'' in the northwest.

Two distinct surface brightness edges, which delineate the spiral at a radius of $25''$ in the west and a radius of $14''$ in the east, are readily seen (Figure~\ref{fig.xraycore}). 
The western edge is very sharp at position angles between $320\degr-370\degr$, but blurs further northward and southward. 
The eastern edge is prominent in the direction between $140\degr-180\degr$, but loses its clear trace southward as it merges into the ``rim", i.e., the innermost core.

We produced a residual image (right panel in Figure~\ref{fig.xraycore}), using \href{http://www.iasf-milano.inaf.it/~eckert/newsite/Proffit.html}{Proffit},
 a custom surface brightness fitting software \citep{eckert2011}. 
To do so, we generated the radial surface brightness profile centered on the X-ray peak ([R.A., DEC]=[16:32:47.0,+5:34:31]), 
 fitted the profile with a one-dimensional $\beta$-model \citep{cava76}, and subtracted the model image. 
 The residual image shows a bright core corresponding to the rim, with a small depression immediately northwest to it. 
There is an excess at the position of the western edge, and a depression immediately outside the eastern edge (between $14''-25''$), 
which is understandable because the western edge increases the overall model surface brightness at those positions. 
There is also an excess corresponding to the tail.


\subsection{Temperature and density profiles across the cold fronts}

\begin{figure*}[hbt!]
\centerline{
\includegraphics[width=1.\textwidth]{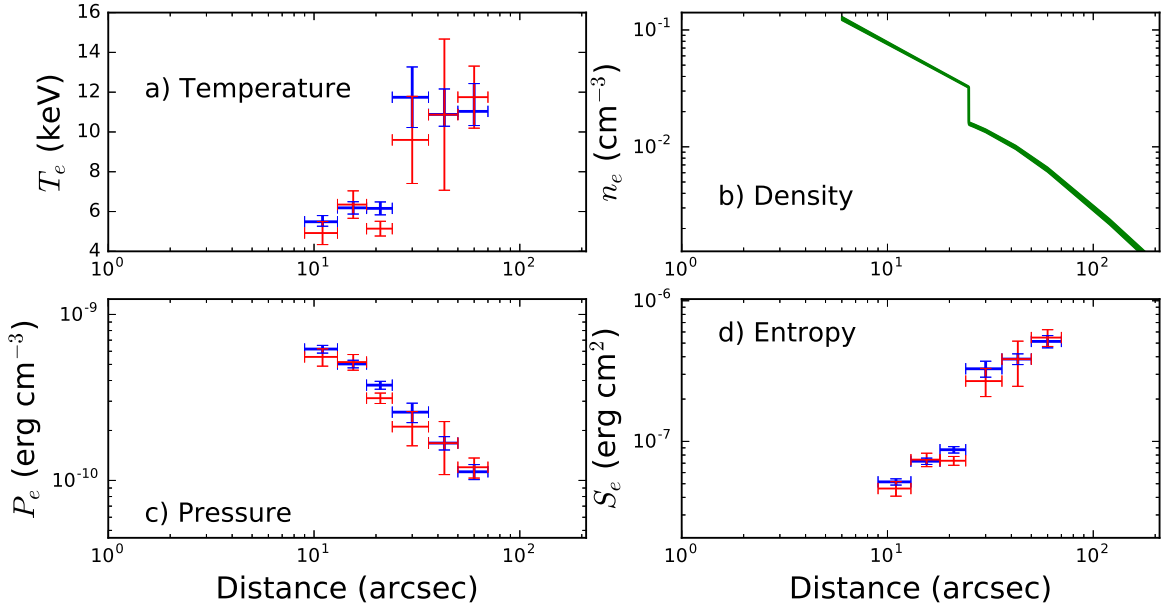}
}
\caption{
Gas properties of the western edge. (a) Temperature, blue data are obtained by fitting an absorbed APEC model, red data are obtained by first
deprojection by dsdeproj and then fit by the same model; (b) Electron density; (c) Pressure $P_e = n_ek_{\rm B}T$, 
(d) Entropy $S_e= k_{\rm B}T n_e^{-2/3}$. 
Blue and red data in (c) and (d) are combined using blue and red temperature data in (a) and the density of each center bins in (b) respectively.
\label{fig:westplot}
}
\end{figure*}

\begin{figure*}[hbt!]
\centerline{
\includegraphics[width=1.\textwidth]{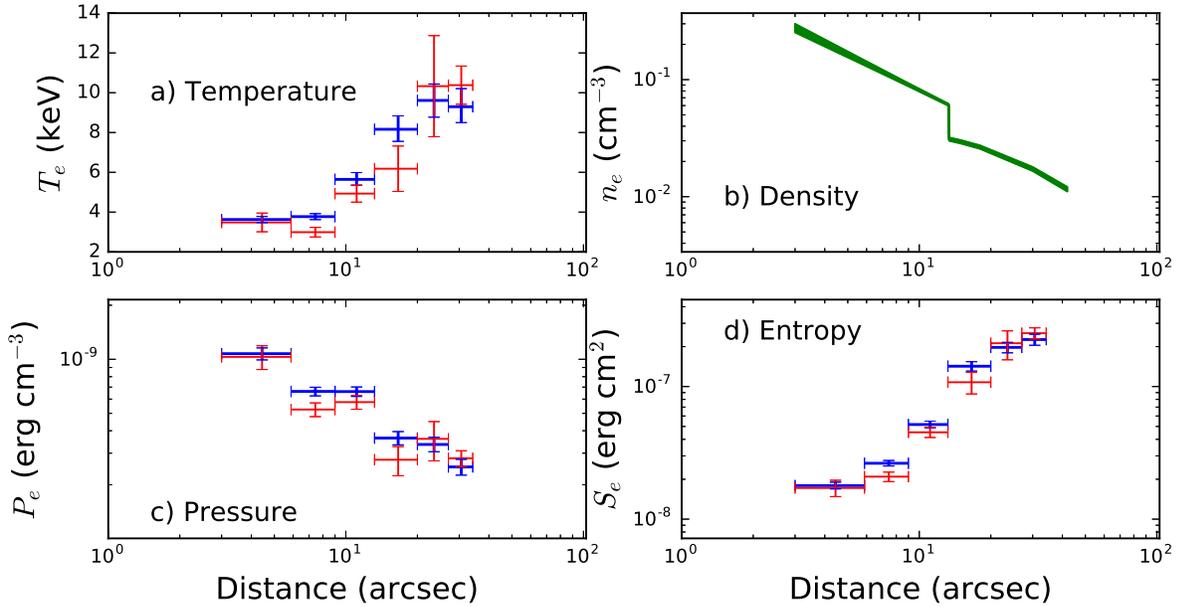}
}
\caption{
Similar to Figure \ref{fig:westplot}, but for the eastern edge.
\label{fig:eastplot}
}
\end{figure*}

The morphology and location of the two edges strongly suggest that they are the result of gas sloshing due to a recent gravitational perturbation of the ICM. 
In the Appendix we discuss the perturber candidates.
Previous work \citep{sanders2005} have shown that the two edges are CFs.
Here using the deeper exposure, we perform an in-depth study of their physical properties.
For each of the two edges, we extract spectra from six concentric sectors (Figure~3b) and fit them with the wabs*apec model to obtain the projected temperature profiles across the CFs.
We can see that the temperature jump at the western CF is as large as by a factor of $\sim$2 (Figure~\ref{fig:westplot}a).
The temperature jump at the eastern CF is not as abrupt as at the western CF (Figure~\ref{fig:eastplot}a). 
We also used {\it dsdeproj} \citep{sanders2007,russell2008}, which assumes spherical symmetry, to obtain the deprojected temperature profile  (red points in Figures~\ref{fig:westplot} and \ref{fig:eastplot}). 
This method subtracts photons emitted from the outer region, therefore leading to a lower 
temperature in the inner regions, with larger error bars.
It is noteworthy that since the cluster core is sloshing predominantly in a plane, 
the spherical symmetry assumed by this method is not exact and is a source of systematic error.
Nevertheless, the deprojected temperature profiles at both edges support the conclusion that they are CFs.

Next, we characterize the gas density profile, for which we adopt the following form,
\begin{equation}
    \begin{split}
        n_e(r)=n_{e0}&
             \left\{
                \begin{array}{lcl}
                    J[1+({r_{j}\over r_{c}})^2]^{-3\beta/2}({r\over r_{j}})^{-\alpha},  & r<r_{j}\\
                    {[1+({r \over r_c})^2]}^{-3\beta/2}, & r~{\geq}~r_{j}\\
                \end{array}
			\right .
    \end{split}
\end{equation}
to account for the density jump ($J$) across the edge (at the break radius $r_j$). 
Here, the density profile is described by a power-law with slope $\alpha$ and a $\beta$-model inside and outside $r_j$, respectively.

The above parameters, except $n_{e0}$, are constrained by fitting the surface brightness profile across the two CFs. 
The western CF is well fitted by this model ($\chi^2$/d.o.f. = 109.4/110), and the very small error ($\sim 0\farcs1$) on $r_j$ indicates that the position of the edge is well-determined. 
To test this, we artificially increase $r_j$ by a value of 0\farcs5 (i.e., the ACIS pixel size) greater than the best-fit value (24\farcs8), and redo the fit fixing $r_j$. This results in $\chi^2$/d.o.f. = 139.6/111. 

The fit to the eastern CF is poorer ($\chi^2$/d.o.f. = 162.9/124), which is apparently due to the variation in the break radius with the azimuthal angle, while our fit necessarily assumes the same $r_j$ within the fitted range of $120\degr-180\degr$. 
Therefore, we further divide the eastern CF into three smaller sectors ($120\degr-140\degr$, $140\degr-160\degr$ and $160\degr-180\degr$) and redo the fit.  
The resultant $r_j$ varies from 12\farcs9 to 14\farcs1, with significantly improved reduced $\chi^2$ (1.08, 1.05 and 1.21).

Finally, we determine the normalization parameter, $n_{e0}$, in Eqn.~1, relating it to the {\it emission measure}, which is derived from the APEC model fitted to the spectrum of the sector immediately outside each CF (the fourth bin in Figures \ref{fig:westplot} and \ref{fig:eastplot}). By definition of the emission measure ($EM$), we have

\begin{equation}
{\small
\begin{aligned}
EM & = \int n_e n_H dV \\
& = {n_{e0}^2 \over n_{e}/n_{H}} r_c^3\theta_{r}\tilde{B}(3\beta-1/2,1/2)\int_{\frac{b_l}{r_c}}^{\frac{b_h}{r_c}} b(1+b^2)^{-3\beta+1/2}db,
\end{aligned}
}
\end{equation}
where $\theta_{r}$ is the opening angle of the CF, and $b_h$, $b_l$ the outer and inner radii of the sector. 
$\tilde{B}$ is the beta function, obtained by integrating density square along the line-of-sight.
 The APEC norms (equivalent to $EM$) are $(4.34\pm0.15) \times10^{-4}~\rm~cm^{-5}$ and $(4.34\pm0.10) \times 10^{-4}~\rm~cm^{-5}$ for western and eastern regions, respectively. The calculated densities are listed in Table \ref{tb.gasprop}.

We also calculated the pressure $P_e=n_ek_{\rm B}T$ and entropy $S_e=k_{\rm B}Tn_e^{-2/3}$.
The density at the temperature bin center is used to compute the pressure and entropy. 
The errors in density were calculated using a Monte Carlo method, assuming Gaussian errors in the fitting parameters.
The errors in pressure and entropy were calculated in a similar way.
Blue data are calculated using projected temperature and red data are calculated using temperature obtained by dsdeproj.
As shown in Figure \ref{fig:westplot}, the western edge has a continuous pressure profile and an abrupt entropy jump.
The pressure profile at the eastern edge is not as continuous as in the western edge. 

\begin{deluxetable*}{cccccccc}
\tablewidth{0.9\textwidth}
\tablecaption{Fitted parameters of the surface brightness profiles across the edges\label{sb.par}}
\tablehead{\colhead{Region}&\colhead{Azimuthal range}&\colhead{$\alpha$}&\colhead{$r_c$ (arcmin)}&\colhead{$\beta$}&\colhead{$r_{j}$ (arcmin)}&\colhead{$J$}&\colhead{$\chi^2$ (d.o.f)}}\\
\startdata
West& $320\degr-10\degr$ &  $0.953^{+0.021}_{-0.021}$&  $0.506^{+0.023}_{-0.023}$   &$0.561^{+0.006}_{-0.006}$& $0.414^{+0.002}_{-0.002}$&  $2.05^{+0.05}_{-0.05}$ &    109.391 (110)\\
\\
East& $120\degr-180\degr$ &  $1.013^{+0.055}_{-0.055}$&  $0.393^{+0.013}_{-0.012}$  &$0.570^{+0.006}_{-0.006}$& $0.222^{+0.001}_{-0.001}$&$1.96^{+0.05}_{-0.05}$ &  162.877 (124) \\

\enddata
\tablecomments{The fitting radial range is 0\farcm1-4\farcm0 for the western edge and 
0\farcm1-2\farcm0 for the eastern edge.}
\end{deluxetable*}

\begin{deluxetable*}{ccccc}
\tablecaption{Gas properties across the edges \label{tb.gasprop}}
\tablehead{\colhead{Region}&\colhead{$n_{\rm in}$}&\colhead{$n_{\rm out}$}&\colhead{$T_{\rm in}$ (dsdeproj)}&\colhead{$T_{\rm out}$ (dsdeproj)} \\
\colhead{} & \colhead{cm$^{-3}$} & \colhead{cm$^{-3}$} & \colhead{keV}  & \colhead{keV}  }\\
\startdata
West&   $0.032^{+0.001}_{-0.001}$&  $0.0158^{+0.0004}_{-0.0004}$ & $6.16^{+0.32}_{-0.32} (5.14^{+0.41}_{-0.39})$ & $11.74^{+1.53}_{-1.52} (9.60^{+3.63}_{-1.89})$ \\
\\
East&   $0.061^{+0.002}_{-0.002}$&  $0.0310^{+0.0006}_{-0.0006}$  &$5.64^{+0.34}_{-0.30} (4.93^{+0.44}_{-0.42})$& $8.16^{+0.68}_{-0.61} (6.18^{+1.32}_{-1.01})$ \\
\enddata
\tablecomments{Density and temperature immediately inside and outside the edges.}
\end{deluxetable*}

\section{Discussion: Suppression of Kelvin-Helmholtz Instability}

\subsection{Effects of gravity and magnetic fields}

The lack of significant KHI at the cold fronts suggests that they are suppressed by some physical process. 
The simplest possible explanation is that the fronts are stabilized against KHI by gravitational acceleration. 
The condition for stability against sinusoidal perturbations at the cold front interface is \citep{vikhlinin2002}:
\begin{equation}
\frac{g}{k} > (v_{\rm in}-v_{\rm out})^2\frac{\rho_{\rm in}\rho_{\rm out}}{\rho_{\rm in}^2-\rho_{\rm out}^2}
\end{equation}
where $\rho$ is the gas mass density, $v$ the velocity tangential to the front surface, $g$ the magnitude of the gravitational acceleration, and $k$ the wavenumber. 
This equation implies that there is a minimum wavelength $\lambda_{\rm min} = 2\pi/k_{\rm max}$ below which KHI perturbations are not suppressed by gravity.

For this calculation, we may use the measured values of the densities $\rho_{\rm in}$ and $\rho_{\rm out}$ from Table \ref{tb.gasprop},
 but since we are unable to directly measure the velocity of the gas above and below the front surface, 
 we must use estimates of the velocity shear in such systems from simulations and theoretical arguments.
 \citet{keshet2012} formulated a theoretical model for spiral sloshing flows in cluster cool cores, 
 arguing that the overall pattern of motion can be separated into a cold, ``fast'' flow and a hot, ``slow'' flow, 
 implying that $|v_{\rm in}| \gg |v_{\rm out}|$, which is also seen in simulations \citep{ascasibar2006, zuhone2010, roediger2011, roediger2012}. 
 Simulations also show that the cold flow is subsonic, with typical Mach numbers of ${\cal M}_{\rm in} \sim 0.3-0.5$ \citep{ascasibar2006, roediger2011}. 
 Given this fact from the simulations and the measured value of the sound speed $c_{\rm s,in}$ underneath the front surface, 
 we may make a rough estimate for the tangential velocity of the cold gas $v_{\rm in}$, which we show in Table \ref{tab:KHI_supp}. 
 
Determining the gravitational acceleration $g$ from the X-ray data is not straightforward, since the gas in the core is not in an equilibrium state, as evidenced by the sloshing motions, and hydrostatic equilibrium cannot be assumed in this region. However, N-body/hydrodynamics simulations of sloshing (Ascasibar \& Markevitch 2006, ZuHone, Markevitch \& Johnson 2010) indicate that the DM core of the main cluster should be relatively stable against the influence of the subcluster, and that hydrostatic equilibrium is a good approximation outside the sloshing region. These two facts together indicate that fitting a mass profile to these outer regions and extrapolating it inward should give a reasonable estimate of the mass inside the CFs.

For this purpose, we use the deprojected density and temperature profiles in the radial range of $r \sim 100 - 1000$~kpc to compute a mass profile assuming hydrostatic equilibrium, and fit this profile to an NFW profile (Navarro et al. 1996):
\begin{equation}
\rho_{DM}(r)={{\rho_s}\over { {{r}\over{R_s}} (1+ {{r}\over{R_s}})^2}}.
\end{equation}	
The parameters we obtain are $\rho_s=4.1\pm0.4\times10^6 M_\odot { \rm~kpc^{-3} }$ and $R_s=256\pm13 \rm~kpc$. 
The implied gravitational accelerations at the western and eastern CFs
are $6.8\pm 1.4 \times 10^{-8}\rm~cm~s^{-2}$ and $7.8\pm 1.5 \times 10^{-8} \rm~cm~s^{-2} $, respectively.

 In Table \ref{tab:KHI_supp} we show the minimum wavelength $\lambda_{\rm min}$ of KHI modes that can be suppressed by gravity. 
 At either the western or eastern front, $\lambda_{\min}$ is still quite large compared to the size of the cold fronts (despite the uncertainties), 
 indicating that gravity is incapable of suppressing perturbations that should still be visible.

We note that Richard et al. (2010) used strong gravitational lensing to derive the total mass in the cluster core, which we find to be nearly twice our derived mass at a radius of $\sim$65 kpc. In this case, the minimum wavelength 
that can be stabilized becomes half that quoted in Table \ref{tab:KHI_supp}.

One must also consider the possibility that although the cold front discontinuity is indeed unstable to KHI, there simply has not been enough time for perturbations to grow. To determine whether this is the case, we calculate the KHI growth timescale via
\begin{equation}
t_{\rm KH} = \frac{2\pi}{\omega_{\rm KH}} = \frac{\rho_{\rm in}+\rho_{\rm out}}{\sqrt{\rho_{\rm in}\rho_{\rm out}}}\frac{\lambda}{|v_{\rm in}-v_{\rm out}|}
\end{equation}
and tabulate the timescales $t_{\rm KH}$ corresponding to $\lambda_{\rm min}$ in Table \ref{tab:KHI_supp}. 
The timescales are rather long (with substantial uncertainty), indicating that for perturbations at these wavelengths, 
they may simply have not had enough time to grow since the formation of the cold fronts.  
However, perturbations at smaller wavelengths will grow faster. For example, a perturbation with a wavelength of 10~kpc will grow with a timescale of $t_{\rm KH} \sim 45$~Myr for both CFs, so perturbations of these scales may be expected to be observed.

\renewcommand{\arraystretch}{1.5}
\begin{table*}
\tabletypesize{\scriptsize}
\caption{Derived Properties of the Cold Fronts\label{tab:KHI_supp}}
\begin{center}
\begin{tabular}{cccccc}
\hline
\hline
Region & $c_{\rm s,in}$ (km/s) & $v_{\rm in}$ (km/s) & $\lambda_{\rm min}$ (kpc) & $t_{\rm KH}$ (Myr) & $B_{\rm in}$ ($\mu$G) \\
\hline
West & 1180$\pm$50 & 470$\pm$120 & 43$\pm$18 & 190$\pm$90 &  24$\pm$6 \\
East & 1150$\pm$50 & 460$\pm$120 & 38$\pm$16 & 170$\pm$80 &  32$\pm$8 \\
\hline
\end{tabular}
\item \small {\bf Note.} --- Sound speeds are calculated from deprojected temperature.
\end{center}
\end{table*}

This indicates that such perturbations may be suppressed by the action of a strong magnetic field layer aligned with the front surface. Such layers should arise naturally at sloshing cold fronts \citep{keshet2010,zuhone2011}, which are produced by the shear amplification of an existing magnetic field, stretching the magnetic field lines parallel to the cold front surface and increasing the magnetic field strength. The minimum magnetic field in the layer necessary to suppress KHI is given by \citep{vikhlinin2002,keshet2010}:
\begin{equation}
B_{\rm in}^2+B_{\rm out}^2 > 4\pi\frac{\rho_{\rm in}\rho_{\rm out}}{\rho_{\rm in}+\rho_{\rm out}}(v_{\rm in}-v_{\rm out})^2
\end{equation}
where $B_{\rm in,out}$ are the magnetic field strengths on either side of the front. The simulations of \citet{zuhone2011} showed that sloshing motions amplify the magnetic field strength predominantly {\it below} the front, so $B_{\rm in}^2+B_{\rm out}^2 \approx B_{\rm in}^2$, giving an estimate for the minimum magnetic field strength in the cold flow. These estimated magnetic field strengths, on the order of 10 ${\mu}$G, are listed in Table \ref{tab:KHI_supp}. Such field strengths could be easily produced by shear amplification of the magnetic field at the cold front surfaces, as demonstrated in simulations by \citet{zuhone2011}.

\subsection{Estimate of the Velocity Difference Across the Cold Front}\label{sec:vdiff_est}

In the previous subsection, we drew on the results of simulations to infer the probable velocity of the sloshing cold gas underneath the cold fronts. 
To the best of our knowledge, with the exception of Keshet et al. 2010, no works have attempted to measure the tangential velocity underneath a sloshing cold front. Previous works (e.g., Vikhlinin et al. 2001) have attempted to estimate the velocity in the free-stream region ahead of a merger cold front, but the sloshing scenario here is distinct from a major merger.
Therefore it is worth making the velocity estimate to see if it is consistent with what we expect from simulations.

\citet{markevitch2001} noted an unphysical jump in the mass profile derived from the X-ray data within an angular sector containing a cold front edge in the relaxed cluster A1795. In such a case, although the gas pressure itself is continuous, its first derivative is discontinuous. They argued that this was evidence for radial acceleration of the gas underneath the front. \citet{keshet2010} later noted that this would cause an unphysical gap to open along the cold front. Instead, they formulated an argument that the difference in gravitational acceleration is compensated by the centripetal acceleration of the tangentially moving gas underneath the cold front surface. Under suitable assumptions, they derived the following expression for the velocity shear (their Equation 2):
\begin{equation}
\frac{\Delta{v^2}}{r} = -\Delta{a_r} = \Delta\left[\frac{1}{\rho}\frac{\partial{P}}{\partial{r}}\right]
\end{equation}
where $\Delta{v^2} = v_{\rm in}^2 - v_{\rm out}^2 \approx v_{\rm in}^2$. We will use this expression to make an estimate for the velocity $v_{\rm in}$ of the cold component of the western cold front, since it exhibits the properties of pressure continuity with a pressure derivative discontinuity. To compute the pressure derivative, we assume that, in the immediate vicinity of the cold front, the temperature is constant with radius on both sides and that the behavior of the density on either side of the front is given by the $\beta$-model and power-law fits from above (Equation 1). 
We further assume for simplicity that the motion of the cold front is in the plane of the sky, which is not unreasonable according to its spiral-like morphology. We determine that $v_{\rm in} = 510\pm320$~km~s$^{-1}$, corresponding to a Mach number of ${\cal M}_{\rm in} = v_{\rm in}/c_{\rm s,in} = 0.43\pm0.27$, which is very similar to the value taken from the simulations discussed in Section 5.1. We note that the value we obtain for the dimensionless pressure discontinuity parameter from \citet{keshet2010} $\delta = (r/c_{\rm s,in}^2)\Delta{a_r} = -0.19\pm0.23$ is marginally consistent with their upper limits for this parameter for the western cold front of A2204 (see their Figure 2).

The relatively large error bars on our estimates for the velocity of the gas underneath the front and the pressure discontinuity point to the need for further study. Despite the fact that our deep exposure of A2204 and the apparent smoothness of its western cold front provides the opportunity to attempt such an estimate, it is still difficult to measure this number accurately, primarily due to the uncertainty in the gas temperature on either side of the front. Additionally, this estimate depends on the assumptions that the gas moving underneath the front can be approximated as having circular motion and that this motion is in the plane of the sky. The best way to quantify the uncertainty associated with these assumptions would be to analyze mock X-ray observations of simulations of cold fronts, where we could carry out the same analysis as for A2204, but would be able to project the cold front along slightly different lines of sight and compare our estimate for the velocity directly to the velocity field in the simulation. We leave this for future work.

\section{summary}
We present an analysis of the combined {\it Chandra} observation of the rich cluster A2204.
This deeper dataset allows us to study the substructures in the ICM in more detail than previous works.
Based on the X-ray data,
we are able to determine the exact position of the SB edges: The western edge is located at $25''~(\approx65~\rm~kpc)$ 
from the X-ray peak, and it is most prominent within the direction $-40\degr - 10\degr$. 
From the direction of $30\degr$ further north and $-40\degr$ further south, the edge is no longer clear. 
At the western edge, the projected temperature profile has a huge jump ($T_{out}/T_{in}=1.91\pm0.27$) and the density has a prominent drop ($n_{\rm{in}}/n_{\rm{out}}=2.05\pm0.05$), while the pressure profile is continuous. Therefore, it is a typical cold front.
The eastern edge is closer to the center (at the radius of $13''\approx35 ~\rm~kpc$), and it is prominent within the direction of $120\degr - 180\degr$, especially in the middle. At this edge $T_{\rm{out}}/T_{\rm{in}}=1.45\pm0.15$ and $n_{\rm{in}}/n_{\rm{out}}=1.96\pm0.05$. Thus the east edge is also a typical cold front.

Both CFs show no visible Kelvin-Helmholtz eddies in {\it Chandra}'s X-ray images.
This suggests that KHI is strongly suppressed on these fronts. 
We have investigated whether or not gravity could suppress KHI, and found that the gravitational acceleration of the underlying DM halo is likely too weak to provide adequate suppression for all wavelengths that would be visible ($\lambda_{min} < 50\rm~kpc$). 
It may be the case that there has not simply been enough time for these perturbations to develop, but this is also unlikely for smaller-wavelength perturbations.
 Therefore, we have investigated the possibility that a magnetic field may be responsible for KHI suppression. 
Based on the front properties, we estimate that the minimum value of the magnetic field to suppress KHI is $ 24 \pm 6{\rm~\mu G}$
 and ${ 32\pm 8{\rm~\mu G}}$ for the western and eastern CFs, respectively. 
Finally, we also use the pressure derivative discontinuity at the western CF to make an estimate for the tangential velocity of the gas underneath the front, and find that the gas is moving with $M_{\rm{in}} = 0.43 \pm 0.27$, consistent with the velocity of sloshing gas seen in simulations.

\vskip 0.5cm
This work was supported by NASA grant GO3-14131X and NSFC grants J1210039 and 11133001.
H.C. is grateful to the warm hospitality of the Center for Astrophysics during her summer visit, and acknowledges support from Top-notch Academic Programs Project of Jiangsu Higher Education Institutions. Z.L. acknowledges support from the Recruitment Program of Global Youth Experts.

%
%
%

\appendix
\section{Candidates of the perturber}
\setcounter{figure}{0}
\renewcommand\thefigure{A\arabic{figure}}
The spiral-like sloshing feature indicates that the orbital plane of the perturber be roughly perpendicular to the line-of-sight \citep{ascasibar2006}.
We are thus guided to look for candidates for a perturbing subcluster. 

\begin{figure}[hbt!]
\centerline{
\includegraphics[width=0.5\textwidth]{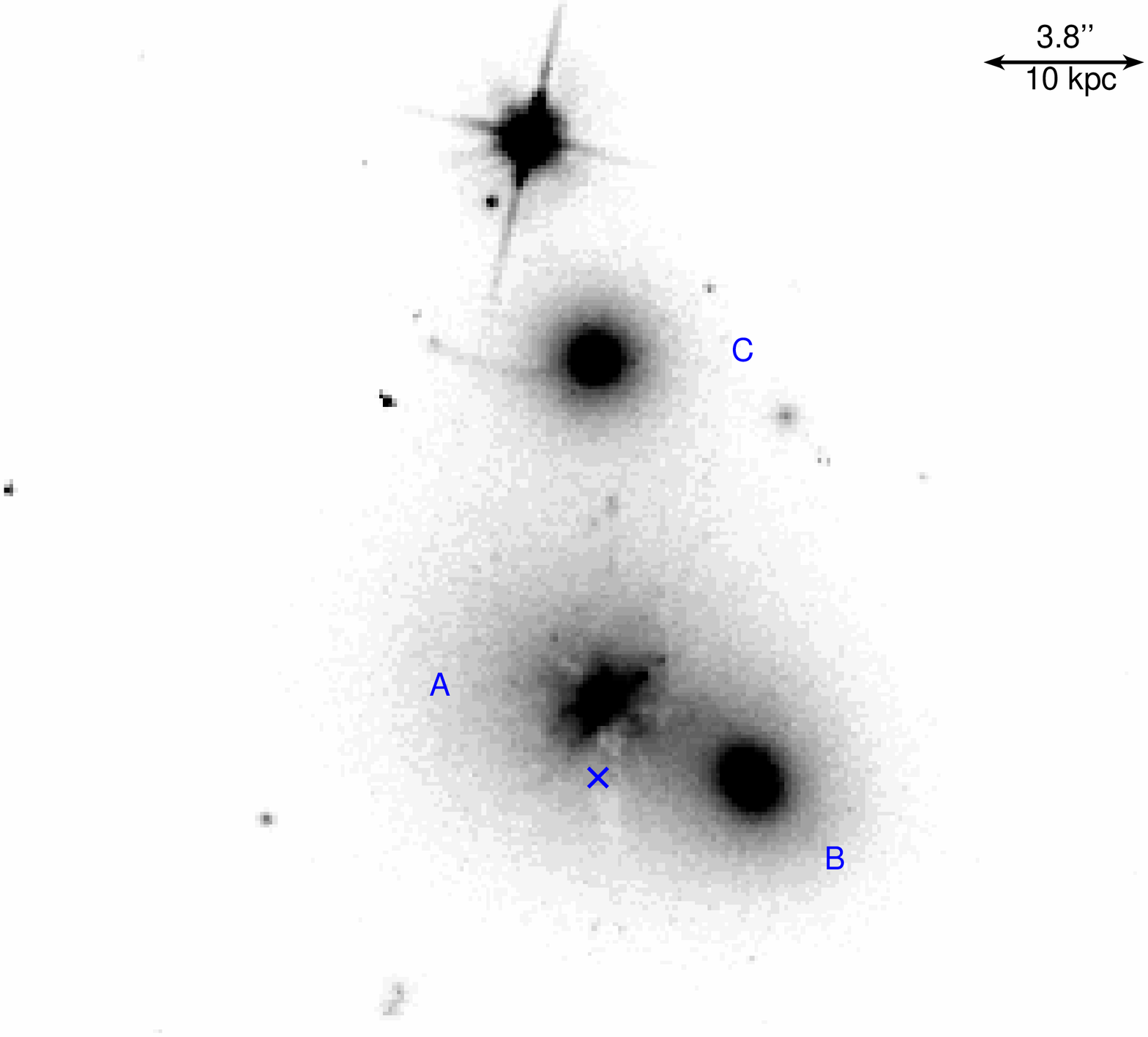}
}
\caption{
HST/WFPC2/F606W image of the central region in A2204. The two central galaxies (A and B) are probably interacting with each other. Galaxy C in the north has a redshift
of z=0.139, or a radial velocity of $3.9\times10^3\rm~km~s^{-1}$ relative to galaxy A, which appears high for a member galaxy. 
The arc near C is a lensed background galaxy \citep{wilman2006}.
\label{fig.hst}
}
\end{figure}

In the core, three large galaxies can be identified in Hubble Space Telescope (HST) Wide-Field Planetary Camera 2 images (Figure~\ref{fig.hst}; \citealt{wilman2006}). 
The galaxy labeled ``A'' in Figure~\ref{fig.hst} is the central cluster galaxy (CCG) of A2204,
which is probably interacting with the neighboring galaxy ``B" \citep{wilman2006}.
In the far-ultraviolet, galaxy ``A'' shows filamentary structures, indicating active star formation \citep{oonk2011}. 
The X-ray peak has a small ($\sim$2$"$) offset to the south from the center of the CCG, which can be supporting evidence of gravitational perturbation of the ICM. 
On the other hand, interaction between CCGs is unlikely to result in the spiral structure.
An alternative scenario is that galaxy ``B" is the remnant of a subcluster that has just finished its second passage, which would have caused gas sloshing at the first passage long time ago \citep{johnson2012}. 
This picture, however, is not supported by the regular morphology of the two edges and their relatively small distances from the cluster center \citep{zuhone2011}.

Because a perturbing subcluster spends most of the time in the cluster outskirts \citep{ascasibar2006},
we searched for the perturber in the $8''\times8''$-binned image (covering a few Mpc across A2204), and found two interesting extended sources, one located 
in the southeast and the other, fainter one in the east (Figure~\ref{fig.bin16}).
We fitted the spectra of the southeast source 
and found its temperature to be $2.73\pm0.42 \rm~keV$, assuming the same redshift as A2204.
Its 0.5-2 keV luminosity is $3.22^{+0.18}_{-0.19} \times10^{43}\rm~erg s^{-1}$.
Using the scaling relation in \citet{zhang2011}, this leads to a gas mass of $\approx 2\times10^{13} M_\odot$
and a dynamical mass of $\approx 2\times 10^{14} M_\odot$, assuming that the ICM accounts for $10\%$ of the total mass.
This appears to be a reasonable mass for a perturbing subcluster. 

\begin{figure}[hbt!]
\centerline{
\includegraphics[width=0.5\textwidth]{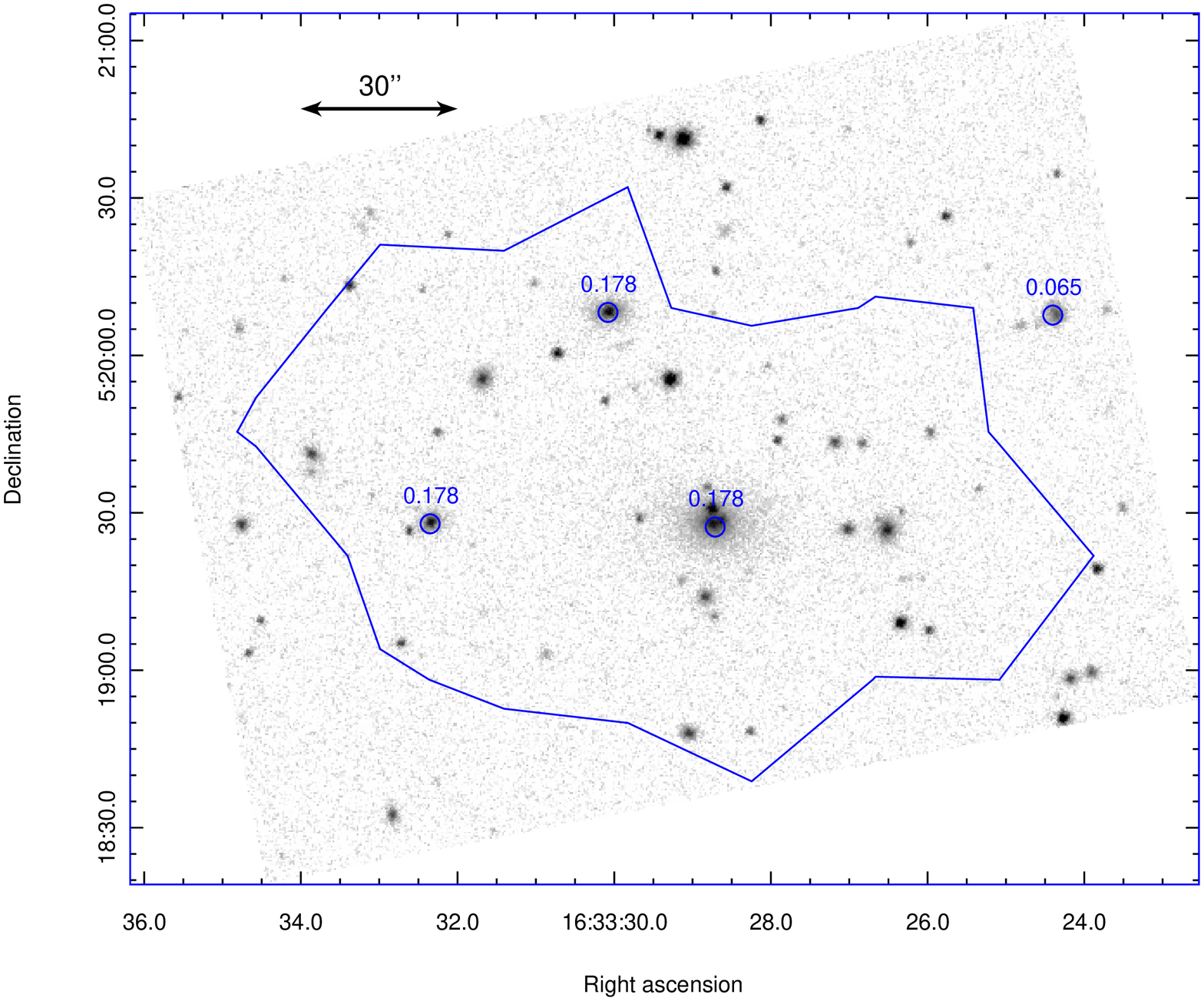}
}
\vspace{-0.3cm}
\caption{
Sloan Digital Sky Survey r-band image of the possible perturber in the southeast (see Figure~\ref{fig.bin16}),
with an X-ray intensity contour overlaid. The redshifts of the brighter galaxies are marked.
\label{fig.sdssSE}
}
\end{figure}

We then searched for optical counterparts using SDSS images, and found a small group of galaxies clustering
around the southeast object (Figure~\ref{fig.sdssSE}).
However, the mean redshift of these galaxies provided by SIMBAD is 0.178. If these galaxies lie at the distance of A2204,
this leads to a relative radial velocity of $7.8\times 10^3\rm~km~s^{-1}$, uncomfortably high for a subcluster, the orbit of which should be close to 
the plane of the sky to produce the spiral-like sloshing feature.

The extended source in the east has too few X-ray photons for a meaningful spectral analysis. Moreover, no galaxy with a similar redshift as A2204 can be identified around it.
We are thus left with no clear candidate perturber within the {\it Chandra} field-of-view.  

\end{document}